\documentclass[12pt]{iopart}
\usepackage[latin1]{inputenc}
\usepackage[T1]{fontenc}
\usepackage{bm}
\usepackage{graphicx}
\usepackage{amsfonts}
\usepackage{amssymb}
\usepackage{color}
\begin{document}

\title[Semiconductor quantum ring]{Semiconductor quantum ring as a solid-state spin qubit}
\author{El\.zbieta Zipper$^1$} 
\author{Marcin Kurpas$^1$}
\author{Janusz Sadowski$^{2,3}$}
\author{Maciej M Ma\'{s}ka$^1$}
\address{$^1$ Institute of Physics, University of Silesia ul. Uniwersytecka 4, 40-007 Katowice, Poland }
\address{$^2$ MAX-Lab, Lund University, 221 00 Lund, Sweden}
\address{$^3$ Institute of Physics, Polish Academy of Sciences, al. Lotnikow 32/46, 02-668 Warszawa, Poland}
\ead{maciej.maska@us.edu.pl}

\begin{abstract}
The implementation of a spin qubit in a quantum ring occupied by one or a few electrons is proposed. Quantum bit involves the Zeeman sublevels of the highest occupied orbital. 
Such a qubit can be initialized, addressed, manipulated, read out and coherently coupled to other quantum rings. 
An extensive discussion of relaxation and decoherence is presented. 
By analogy with quantum dots, the spin relaxation times due to spin-orbit interaction for experimentally accessible quantum ring architectures are calculated. 
The conditions are formulated under which qubits build on quantum rings can have long relaxation times of the order of seconds. 
Rapidly improving nanofabrication technology have made such ring devices experimentally feasible and thus promising for quantum state engineering. 

\end{abstract}

\maketitle

\section{Introduction}\label{sec1}
Quantum information and computation are one of the fastest expanding areas
of modern physics. Among many different physical implementations of qubits (for a review see, e.g., Ref. \cite{book}) the solid state devices seem to be the most promising because of their scalability, tunability and relatively long coherence times.

The spin of a single electron semiconductor quantum dot (QD) placed in a 
magnetic field ${\bm B}$ is a natural two-level system suitable for use as a qubit \cite{loss1998,vander}. The Zeeman splitting phenomenon is responsible for creation of an energy gap $\Delta _{Z}$ 
between two electron states with opposite spin.
However, quantum confinement properties of QDs can be deeply mutated to
crater-like nanostructures (in other words to ring-shaped QDs) called hereafter quantum rings (QRs) \cite{fuhrer2001,bara,lei,kot,kleemans,mano}.
Just like QDs, 
QRs possess atom-like properties making them attractive
candidates for future device applications in quantum information processing. 
These nanometer-size rings which are the nanoscopic analogues of benzene have many intriguing properties. 
The ability to fill QR with one or a few electrons offers new possibilities, e.g., to detect persistent current (PC) carried by
 single electron states \cite{kleemans} or the magnetoinduced change of the ground state \cite{kot,lei}.

In this paper 
we discuss the possibility of building spin qubits on defect free QRs and show that they can be used for quantum state manipulation. 
Owing to the strong confinement of electrons in QRs the orbital states are strongly quantized and the electron spin states are very stable due to the substantial suppression of spin-flip mechanisms. 
It is well known that QDs are one of the best systems for solid state qubit implementations with relaxation times exceeding seconds. 
It is thus interesting to relate qubits built on QRs to those on QDs. In this context we show that QRs are also attractive for the realization of spin qubits with relaxation times of the same order as for QDs.
The considerations in this paper are general and can be used both for electrostatically defined QRs (EQRs) \cite{fuhrer2001,amasha} and self-assembled QRs (SQRs)\cite{lei,kot,kleemans,mano}. 
EQRs can be primarily controlled electrically, SQRs can be primarily controlled optically. 

In Section \ref{sec2} we introduce basic characteristics of quantum rings,
 discuss the formation of spin qubits and provide a brief description of how to manipulate their states. 
In Section \ref{sec3} we make estimations of the relaxation and decoherence times. General discussion of possible experimental realizations is given in Section \ref{sec4} and conclusions are presented in Section \ref{sec5}.

\section{Quantum confinement of semiconductor quantum ring and the formation of spin qubits} \label{sec2}
Nowadays technology allows the preparation and characterization of very small, high-mobility semiconductor structures
 of dot or ring geometry with very good resolution. Recently several high-quality quantum rings on, e.g., AlGaAs-GaAs heterostructures \cite{fuhrer2001}, InGaAs \cite{kot,lei} and GaAs \cite{mano} have been produced and investigated.
Here we consider 
a semiconductor QR of radius $r_0$ and finite thickness containing a \textit{single or a few electrons}. 
The ring is placed in a static magnetic field parallel ($B_{\|}$ ) or perpendicular ($ B_{\bot}$) to its plane. 
The in-plane orientation is favourable as it does not disturb the orbital levels \cite{hans}. The nanometer size of the ring causes quantum size effects important.

For a $2D$ ring in a static magnetic field $B_{\bot}$ we assume the 
Hamiltonian in the form 
\begin{equation}
H = \frac{1}{2m^{*}}  \left( {\bf p}+ e{\bf A} \right)^2 +\frac{e \hbar}{2m_e } {\bm \hat\sigma  }\cdot {\bm B} + V(r), 
\label{Hamiltonian_pr}
\end{equation}
where $ m^{*}$ is the effective electron mass, $ {\bm A}=(0,xB_z,0)$ is the vector potential, $V(r)$ is the confinement potential the exact form of which will be given later in the text. If the ring is placed in a parallel magnetic field $B_{\|}$ instead of $B_{\bot}$ then $ {\bm A}=0 $.

The energy spectrum of a QR consists of a set of discrete levels $E_{nl}$ due to radial motion with radial quantum numbers $n=0,1,2,\ldots$, and rotational motion with angular momentum quantum numbers $l=0,\pm 1,\pm 2\ldots$.
The single particle wave function is of the form 
\begin{equation}
\Psi_{nl} = R_{nl}\left(r\right)\exp\left(i l \phi \right)\chi_{\sigma},
\label{eq_psi_nl}
\end{equation}
with the radial part $R_{nl}(r)$ and the spin part $\chi_{\sigma}$.
For finite--width QRs both $E_{nl}$ and $\Psi_{nl}$ have to be calculated numerically \cite{chakra}. 
In contradiction to QDs, the energy levels numbered by $n>0$ always lie higher in energy than those with increasing $l$ and they do not enter the following analysis. \\
The application of a magnetic field ${\bf B}$ splits the orbital energy levels by 
\begin{equation}
\Delta _{Z}=g_{s}\mu _{B}B,
\label{delta_z}
\end{equation}
where $ g_s $ is the electron spin $g$-factor and $ \mu_B$ is the Bohr magneton.
Another important energy gap is the distance from the highest occupied orbital state ($l$) to the first excited  orbital state ($l\pm1$),
\begin{equation}
\Delta_{l}=\left\{
\begin{array}{rl} 
E_{0,l \pm 1}- E_{0,l}, & \rm{for } B=B_{\|},\\ 
E_{0,l-1}- E_{0,l}, & \rm{for } B=B_{\bot}.
\end{array} \right.
\label{eq_deltal}
\end{equation}
If the following relation holds

\begin{equation}
k_BT \ll \Delta_Z \ll \Delta_l,
\label{warunek}
\end{equation}
the two Zeeman sublevels of the orbital $l$ are well separated from the others and the ring can be well approximated as a two-state system (a qubit). 
We assume that the \textit{'operating'} orbital $l$ is occupied by a single electron only, i.e., for $l=0$ the number of electrons is $N_e =1$, for $|l|=1$, $N_e =3$, etc. 

In our analysis we consider several different quantum rings. The radii and confining potentials of three of them ($A$,
$B$, $C$) are chosen to roughly reproduce the energy spectra of the recently grown InGaAs rings described in Refs. \cite{lei}, \cite{kot}, and \cite{kleemans}, respectively. The confining potential used in all these cases is assumed to be of the following form,
\begin{equation}
V_1(r) = \frac{1}{2} m^{*}\omega_0^2\left(r - r_0\right)^2,
\label{V_harm}
\end{equation}
where the parameters are collected in Table \ref{table1}. 

\begin{table}
\caption{The parameters and relaxation times of three modelled InGaAs quantum rings corresponding (in alphabetical order) to the experimental rings described in Refs. \cite{lei}, \cite{kot} and \cite{kleemans}. The ring geometry has been reached by the confining potential $V_1(r)$ (Eq. \ref{V_harm}); $ \hbar \omega_0 $ is the potential strength. $B_{\|}=1$T has been assumed.}
\begin{indented}
\item[]\begin{tabular}{@{}ccccccc}
\br
{ Ring} & $ r_0$ [nm] & $ \hbar \omega_0$ [meV] & $ T_1^0$ [s] &$ T_1^1$ [s] & $ T_1^2$ [s] & $ T_1^3$ [s] \\ 
\mr
A & 20 & 15 & 0.015 & 0.053 & 0.067 & 0.071 \\ \mr 
B & 14 & 12 & 0.26 & 0.19 & 0.14 & 0.11\\ \mr 
C & 11.5 & 25 & 1.35 & 1.88 & 1.66 & 1.36 \\ \br 
\end{tabular}
\label{table1}
\end{indented}
\end{table}

\begin{figure}[htb]
\includegraphics[width=\linewidth]{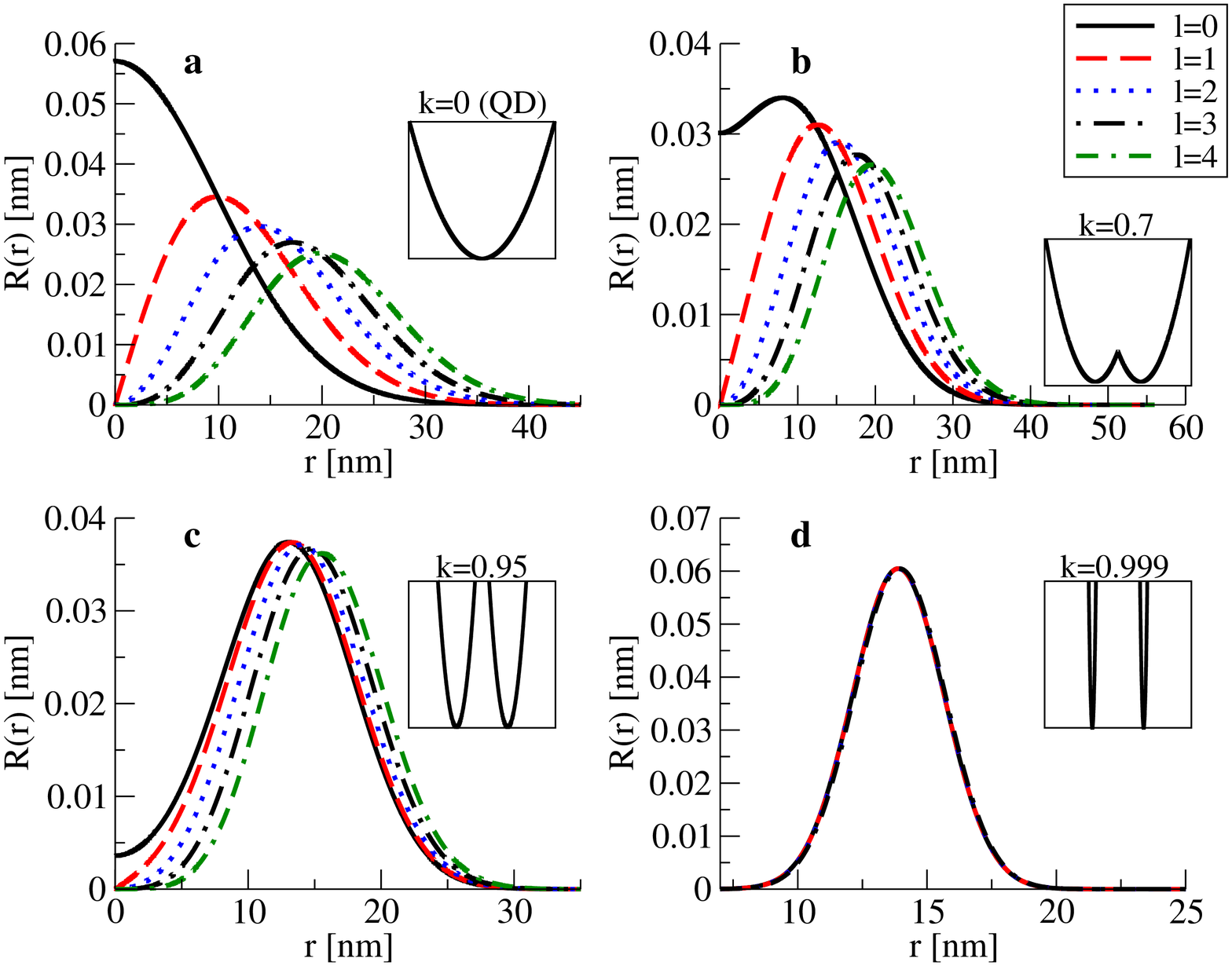}
\caption{The radial part $R_{0l}(r)$ of the electron wave function (\ref{eq_psi_nl}) plotted as a function of radius $r$ for different values the orbital quantum number $l$. Panels $a-d$ include results for different shapes of the confining potential (\ref{V_k}) (shown in the inset plots). In all cases $ r_0=14$nm, $B_{\|}=1$T have been assumed.}
\label{f_radial}
\end{figure}

For the remaining rings we have assumed the same radius as for ring $B$, but the potential takes on different shapes. 
In order to be able to compare results for QRs and QDs, we parametrize the potential in such a way that it can
 reproduce both harmonic potential of a QD as well as a $\delta$--like potential of a quasi one--dimensional (1D) QR.
 It is given by
\begin{eqnarray}
V_2(r)&=&\frac{1}{2} m^{*}\omega_0^2 \left[(1-k)r^2 + \frac{k}{1-k}\left(r-r_0 \right)^2 \right] \nonumber \\
&&=\frac{1}{2}m^{*}\omega_{\rm QD}^2r^2 + \frac{1}{2}m^{*}\omega_{\rm QR}^2\left(r-r_0\right)^2.
\label{V_k}
\end{eqnarray}
It is a superposition of QD and QR potentials, where the confinement (a measure of radial localization of the electron
wave function) is given by $\hbar\omega_{\rm QD}$ and $\hbar\omega_{\rm QR}$, respectively. 
For $k=0$ the second term vanishes and the potential describes harmonic QD. 
On the other hand, in the $k\rightarrow 1$ limit it describes a 1D QR. 
Therefore, changing $k$ from 0 to 1 one can observe how the properties of a quantum system evolve while moving from QD to QR. 
The radius of QR is defined by $r_0$ in Eq. \ref{V_harm}, i.e., it is the distance from the center of the ring to the minimum of the confining potential. 
The definition of the radius of harmonic QD is not so unambiguous -- we use $r_0$ defined by the shape of the ground state wave function $\Psi(r,\phi) \propto \exp(-r/r_0)$. 
In order to ensure that the size of the system does not depend on $k$, and therefore that its properties depend only
on the shape of the potential, in Eq. \ref{V_k} we assume 
\begin{equation}
\omega_0=2\hbar/m^*r_0^2,
\label{omega_0}
\end{equation}
 what gives the radius of the QD equal to $r_0$. Fig. \ref{f_radial} shows the radial part of the wave function for ground
 state ($l=0$) and four lowest excited states ($l=1,\ldots,4$) for different values of $k$:
 for $k=0$ we model QD (Fig. \ref{f_radial}a), for $0<k<1$ we get QRs of decreasing thickness (Fig. \ref{f_radial}b,c), reaching at $k=0.999$ a quasi 1D ring (Fig. \ref{f_radial}d). 
The corresponding shape of the confining potential is shown in the insets.

We would like to stress that our model calculations are for the circularly symmetric nanostructures, whereas some of the experimentally fabricated rings may have slightly different symmetry and therefore slightly different energy spectrum. Additionally, we neglect any imperfectnesses that are present in real rings (impurities, variable thickness, etc.).

In order to use quantum rings in quantum computation it is necessary
to establish a way to perform single qubit operations and to implement
efficient quantum logic gates on pairs of qubits. 
During the past few years a big progress have been made towards full control of quantum states of single and coupled spins in QDs \cite{hans,ladd,aws}. 
Going carefully through all this one finds that most of those features are shared also by ring-shaped QDs.\\
The QR qubit can be initialized by, e.g., thermal equilibration or by optical pumping, coherently manipulated 
(through magnetic resonance technique or by faster electrical and optical gates) and read out using both electrical
and optical techniques \cite{mano,hans,elz}. Coherent coupling of EQRs leading to the formation of, e.g., the CNOT gate can be 
obtained in an analogous way as for QDs \cite{loss1998,vander}, by assembling a system of two coplanar QRs with the
 possibility of tuning their exchange coupling $J$ by gating the barrier between them. 
 Such coupling can be switched on and off by electrical impulses. Quantum gates for SQRs can be accomplished by electronic or photonic connections \cite{mano,ladd,Abba}.
 Single qubit rotations together with the CNOT gate form an universal set of quantum logic gates. Remarkably the operations are very fast, on the order of pico to nanoseconds \cite{przegladLossa}. Thus very many coherent operations can be performed during the decoherence times estimated in Section \ref{sec3}.

Recently a scheme for creating coherent coupling of spin qubits, each placed in a microcavity, by entanglement swapping \cite{kurpas} has been proposed. 
Such long-distance entanglement is a crucial ingredient for quantum communication.

The long term promise of spin qubits depends crucially on the relaxation and decoherence times which are strongly related to the quantization of the orbital states in QRs and can be different than in QDs due to different geometry.

As the experiments were performed mainly at $B\geq 1$T, in the following we fix the magnitude of the magnetic field to $B=1$T. We also assume the electron spin $g$-factor $ \mid g_{s}\mid = 0.8$ for InGaAs samples \cite{kroutvar}, which gives the electron spin Zeeman splitting $\Delta_Z=0.046$meV. 
 
\section{Spin relaxation and decoherence}\label{sec3}

The main difficulty in development of a quantum computer is to keep the qubits in the quantum regime for a sufficiently long time. 
The ideal situation would be to cut off the interaction with the environment that is the main source of destruction of a quantum state. 
This is, however, a very difficult task as this interaction is equally needed (for steering, measurement, etc.) as unwanted (decoherence, relaxation).

Electron spin decoherence is caused primarily by spin-lattice relaxation via phonon scattering and spin-orbit (SOI) interaction and by hyperfine (HFI) interaction with nuclear spins
\cite{Khaetskii,golo,stano}.
At first we discuss the spin relaxation time $T_{1}$. 
At magnetic fields $B < 0.1$T the dominant relaxation mechanism is the HFI but for larger fields this mechanism is suppressed by the mismatch between the nuclear and electron Zeeman energies. 
At $0.5$T$<B<10$T the SOI causes spin relaxation by mixing the spin and orbital states and providing the mechanism for coupling of spins to (mainly) piezoelectric phonons. 
The prolongation of spin relaxation times for small nanostructures stems from a drastic reduction in spin-phonon coupling mediated by the combination of electron-phonon and SOI, due to strong confinement. 

The comprehensive analysis of relaxation due to phonons in QDs has been given in Refs. \cite{Khaetskii}--\cite{geller}. 
It was shown that at $B=1$T the relaxation is dominated by a single--phonon admixture process; it has been verified in several experiments \cite{amasha,hans,kroutvar,imam}.
To discuss the relaxation we have followed Ref. \cite{Khaetskii} and in particular Eq. 7, which is valid for a set of confining potentials and therefore for different shapes of the sample. 
We have used it to estimate relaxation times for circularly symmetric systems discussed above placed in 
the magnetic field $B_{\|}$ ($B_{\bot}$ ). 
We obtained
\begin{equation}
\frac{1}{T_1^l} = 2 C_{ph} \left( \alpha_{xx}^ l\right)^2 \left(1+cos^2 \vartheta\right) \Delta _z^5,
\label{gamma_1}
\end{equation}
where $\vartheta=0$ for $B=B_{\bot}$, $\vartheta=\pi/2$, $B=B_{\|}$; the exact form of $C_{ph}$ is given in Ref. \cite{Khaetskii},

\begin{equation}
\alpha_{xx}^l = \sum_{l''}2e^2 \frac{|<0,l''|x|0,l>|^2}{E_{0,l''}-E_{0,l}} = \frac{4 \pi^2 e^2}{\Delta_{l}} \Xi_{l}^2 ,
\label{alpha_xx}
\end{equation}
where $ 0 $ stands for the quantum number $ n $, $l$ is the orbital number of the highest occupied state, and
\begin{equation}
\Xi_{l} = \int_0^{\infty} R_{0l}^{*}R_{0l'}r^2 dr,
\label{xi}
\end{equation}
is the 'overlap factor', where $ R_{0l}$ is the radial part of the electron wave function (\ref{eq_psi_nl})
and $l'=l-1$ if the field is perpendicular to the ring or $l'=l\pm 1$ for a parallel field.

Inserting Eq. \ref{alpha_xx} into Eq. \ref{gamma_1} and applying little algebra one obtains
the relaxation time for a nanosystem with a single electron in the highest occupied state $l$ given by
\begin{equation}
T_{1}^l = \frac{\eta}{\Delta_ z^5}\;\frac{\Delta_l^2}{ \Xi_{l} ^4},
\label{t1_r}
\end{equation}
where 
\begin{equation}
\eta=\frac{\hbar^5 }{\Lambda_p (2\pi)^4 (m^{*})^2 (1+\cos^2\vartheta)},
\end{equation}
$\Lambda_p$ is the dimensionless constant depending on the strength of the effective spin-piezoelectric phonon coupling and the magnitude of SOI, $\Lambda_p=0.007$ for GaAs type systems \cite{kroutvar,Khaetskii}.

\begin{figure}[htp]
\includegraphics[width=\linewidth]{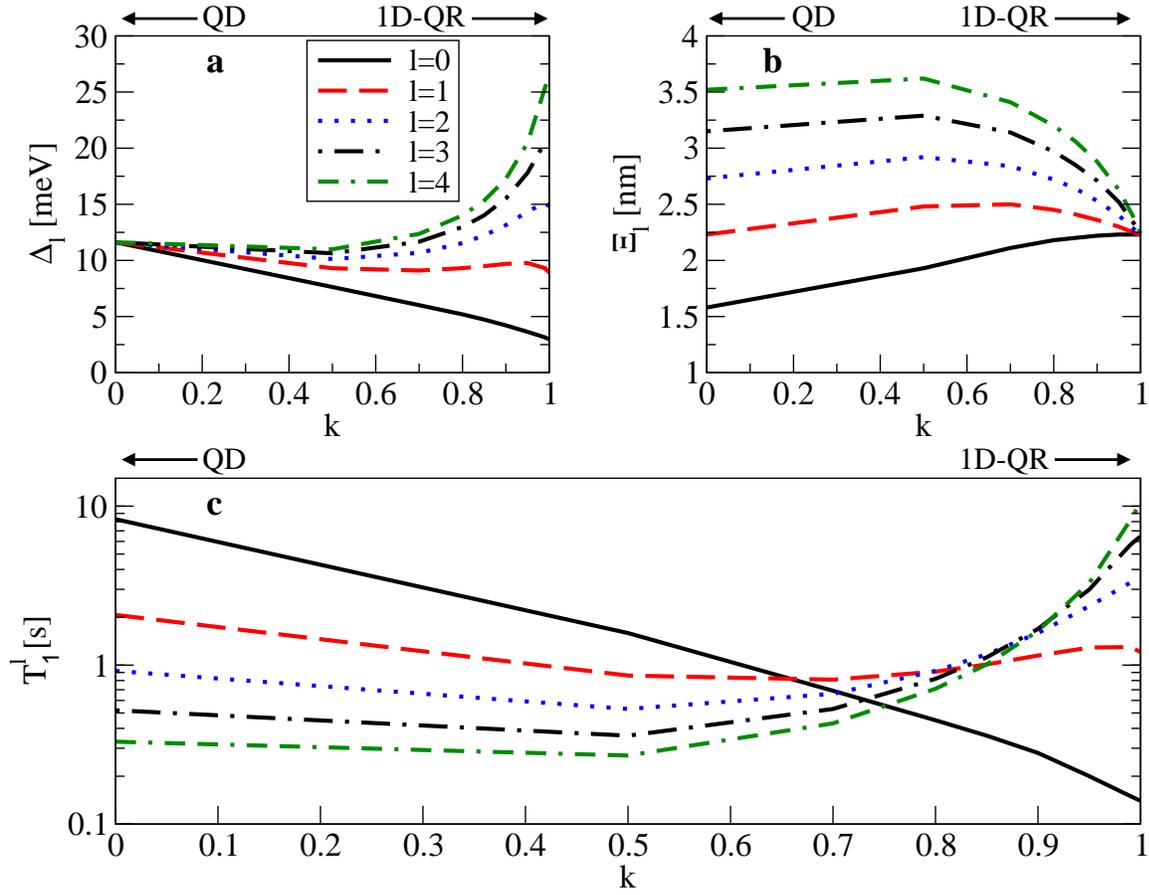}
\caption{a) The orbital energy gap $\Delta_l$ as a function of the potential parameter $ k $, for different values of $ l $ (corresponding to different occupation $N_e$). 
For $ k=0 $ the potential (Eq. \ref{V_k}) models QD and $ \Delta_l $ is $ l $-independent. Increasing $ k $ we reach (for $ k \rightarrow 1$) the 1D-QR limit with $ \Delta_l $ defined by Eq. \ref{delta_R1D}; 
b) The overlap factor $\Xi_l$; c) the relaxation time $ T_1 $ plotted as a function of $ k $ for different orbital states $ l $. $r_0=14$nm has been assumed.}
\label{f_delta_Ksi_T1}
\end{figure}

It follows from Eq. \ref{t1_r} that $T_{1}^l$ depends on $\Delta_l$, i.e., on the number of electrons $N_{e}$
 and on $\Xi_{l}$, i.e., on the wave functions of the neighbouring $l$ states (see Eq. \ref{xi} and 
 Fig. \ref{f_radial}a-d). Notice, that in contrast to QDs where $\Delta_{l}=\hbar\omega_0$, for QRs 
 the energy gaps between neighbouring $l$ states are $l$-dependent and increase with
 increasing $l$ (faster for a thinner ring), tending to 
\begin{equation}
\Delta_{l}^{1D} =\frac{\hbar^{2}}{2m^{*}r^{2}_{0}}\left ( 2 l + 1\right),
\label{delta_R1D}
\end{equation}
for a quasi 1D ring (see Fig. \ref{f_delta_Ksi_T1}a). \\

Let us first discuss the case of a single electron ($N_e=1$, $l=0$) - the results are denoted in Fig. \ref{f_delta_Ksi_T1}a-c as solid black lines.
For QDs ($k=0$) the formula (\ref{xi}) reads 
\begin{equation}
\left(\Xi^{\rm dot}_0\right)^2 = \frac{\hbar^2}{4\pi^2 m^{*} \hbar\omega_0}. 
\label{xi_d}
\end{equation}
Replacing $\Xi_l$ in Eq. \ref{t1_r} by $\Xi^{\rm dot}_0$ gives 
\begin{equation}
T_{1}^{0,\rm dot} = \Lambda_p^{-1} \frac{\hbar \left(\hbar\omega_0\right)^{4}}{\left(\Delta_Z\right)^{5}(1+\cos^2\vartheta)},
\label{t1_r1}
\end{equation}
i.e., the relaxation time for QDs obtained in Ref. \cite{Khaetskii}. 

It is interesting to compare the relaxation times for QRs and QDs -- we show below that $T_{1}^{0,\rm dot}$ is a higher limit of $T^0_{1}$. 
From Eqs. \ref{t1_r} and \ref{t1_r1} one can find the formula relating them: 
\begin{equation}
T_{1}^{0}=T_{1}^{0,\rm dot}\left( \frac{\Delta_0}{\hbar \omega_0}\right)^2\left( \frac{\Xi_0^{\rm dot}}{\Xi_0}\right)^4.
\end{equation}
It follows from our results (Figs. \ref{f_delta_Ksi_T1}a and \ref{f_delta_Ksi_T1}b)
 that for singly occupied nanostructure for arbitrary ring thickness 
 \begin{equation} 
\begin{array}{rl} 
\hbar \omega_0>\Delta_{0}, & \\
 \Xi_{0}^{\rm dot}<\Xi_{0} & 
\end{array} \bigg\rbrace \Rightarrow T^0_1<T_{1}^{0,\rm dot},
\label{D}
\end{equation}
i.e., assuming the same size, QD is the structure having the longest relaxation time. 
To understand the first of the inequalities we compare the formulas for $\Delta_{0}$ for QD (Eq. \ref{omega_0}) and 1D QR (Eq. \ref{delta_R1D}). We see that $\hbar\omega_0= 4\Delta_{0}^{1D}$.
For rings of finite thickness $\Delta_{0}$ changes smoothly between these two values.\newline
The inequality between the overlap factors $\Xi_{0}$ and $\Xi_{0}^{\rm dot}$, follows from the difference in
 shape and distribution of the radial parts $R_{0l}$ (Fig. \ref{f_radial}a-d). 
We see that for QD the radial functions are concentrated closer to $r=0$ than for QRs where they stay mostly at larger $r$. 
Additionally, for QD the difference between the $R_{00}(r)$ and $R_{01}(r)$ is much more significant than for QR. 
Both these properties result in a smaller value of $\Xi_0$ for QD than for QR, leading to the relations (\ref{D}). 

Notice that the relation (\ref{D}) is a consequence of the assumed parabolic shape of the potential confining both the QDs and QRs. For, e.g. a rectangular potential, the charge distribution of the electrons would be different leading to the decrease of the difference between $T_{1}^{0}$ and $T_{1}^{0,\rm dot}$ - the calculations are in progress.
 
\begin{figure}[htp]
\includegraphics[width=\linewidth]{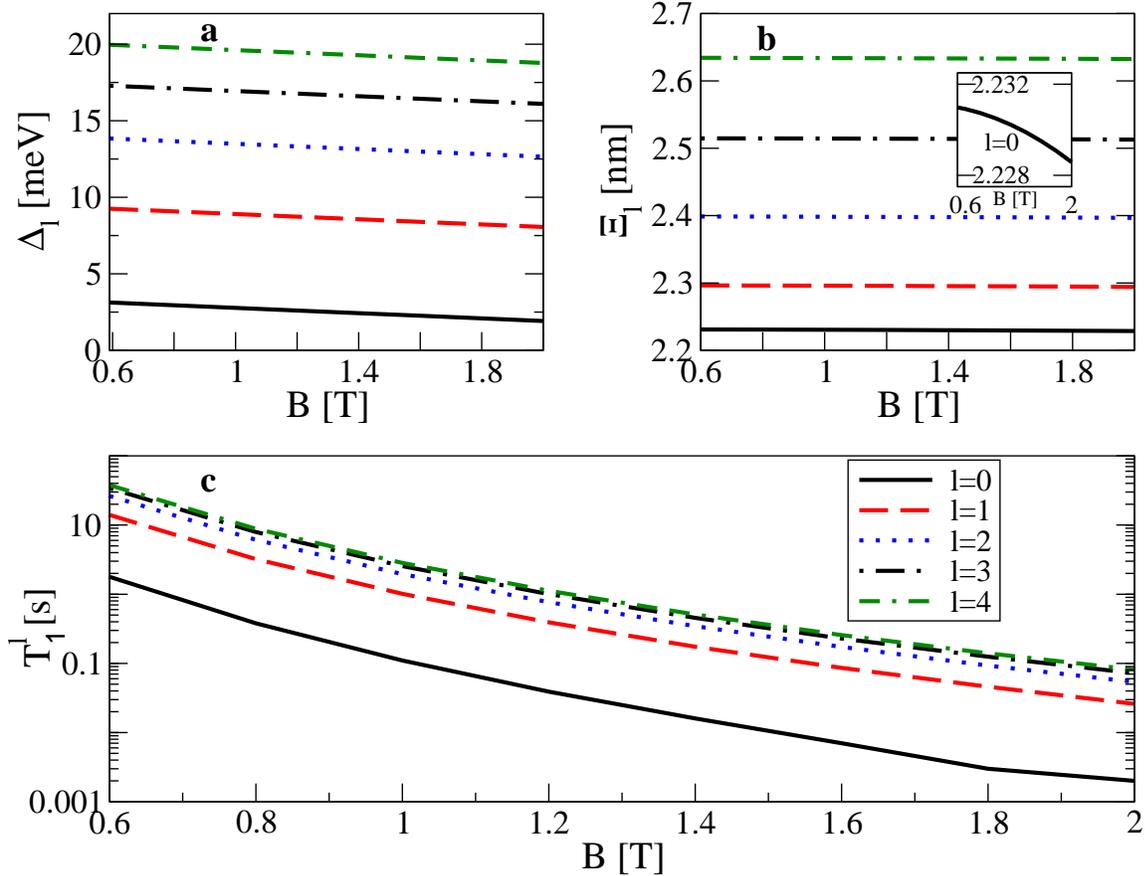}
\caption{a) The orbital energy gap $\Delta_l$ as a function of the magnetic field $ B=B_{\bot} $, for different values of $ l $; b) The overlap factor $\Xi_l$; the inset plot shows the detailed curve $ l=0 $; c) the relaxation time $ T_1^l $. The remaining parameters are: $ k=0.95$, $r_0=14$nm.}
\label{f_delta_Ksi_T1_B_prost}
\end{figure}

The situation, however, changes if the number of electrons is larger than one.
It was shown \cite{fuhrer2001,hans} that for such cases the
electron--electron interaction is well described by the constant interaction
model - it shifts the single electron spectra by a multiple
of the charging energy, $(N_e-1)E_{C}$. Assuming this, the spectra in Figs. \ref{f_delta_Ksi_T1} and \ref{f_delta_Ksi_T1_B_prost} correspond to the energies after the charging energy has been subtracted. 

 One can see from Fig. \ref{f_radial}a that for QDs with $l>0$ the maxima of the wave functions move to larger $r$ leading to an increase of $\Xi_{l}^{\rm dot}$ (Fig. \ref{f_delta_Ksi_T1}b) and subsequent decrease of $T_{1}^{\rm dot}$ (Fig. \ref{f_delta_Ksi_T1}c). 
 For QRs of the large thickness ($0<k<0.8$) the situation is similar as for QDs, but for thinner rings with $k>0.8$ both the decrease of $\Xi_{l}$ (Fig. \ref{f_delta_Ksi_T1}b) and the simultaneous increase 
 of $\Delta_{l}$ (Fig. \ref{f_delta_Ksi_T1}a) lead to a substantial increase of $T_{1}^l$ (see Fig. \ref{f_delta_Ksi_T1}c and Table \ref{table2}). 
 Such relatively thin rings with relaxation times exceeding seconds at $B=1$T are within reach for nowadays nanotechnology.
In Table \ref{table1} we also presented the relaxation times for the rings $ A-C $. We see that they increase considerably with decreasing radius of the rings reaching the value of $T_{1}^0=1.35$s already for singly occupied ring $C$. However because the rings $ A-C $ are relatively thick we do not get the essential increase of 
$T_{1}^l>1$.
\begin{table}
\caption{The relaxation time $T_1^l$ for different values of the orbital number $l$ (equivalently, the number of electrons $N_e$) and different shapes of the potential $V_2(r)$. $r_0=14$nm, $B_{\|}=1$T have been assumed.}
\begin{indented}
\item[]\begin{tabular}{@{}cccccc}
\br
{\bf k}	& \textbf{\emph{l}=0} & \textbf{\emph{l}=1}& \textbf{\emph{l}=2}& \textbf{\emph{l}=3}& \textbf{\emph{l}=4} \\ 
\mr
0&	8.03&	2.01&	0.89&	0.50	&0.32 \\ \mr
0.5	&1.55&	0.83&	0.52&	0.35&	0.26 \\ \mr
0.7	&0.67&	0.78	&0.64&	0.51&	0.41 \\ \mr
0.8	&0.44	&0.88	&0.89&	0.79	&0.69 \\ \mr
0.85&	0.35	&0.98	&1.13&	1.09&	0.99 \\ \mr
0.9&	0.27&	1.11&	1.54	&1.64&	1.59 \\ \mr
{\bf 0.95}	&{\bf 0.20}	&{\bf 1.26}	&{\bf 2.28 }&{\bf 2.91}&{\bf3.19} \\ \mr
0.99&	0.15	&1.26	&3.20&	5.54	&7.92 \\ \mr
1D-QR	&0.13	&1.18&	3.25&	6.28	&10.22 \\ \br
\end{tabular}
\label{table2}
\end{indented}
\end{table}

The results presented in Fig. \ref{f_delta_Ksi_T1} and in Table \ref{table2} have been obtained for the 
magnetic field parallel to the ring. In such a case the movement of electrons is not affected by the field what 
results in a two--fold degeneracy of orbital states $E_{0,l}=E_{0,-l}$. This degeneracy, however, is removed when 
there is a nonzero component of the magnetic field perpendicular to the ring. Then, $E_{0,-|l|}<E_{0,|l|}$ and the 
distance to the first excited state has to be calculated according to the lower line in Eq. \ref{eq_deltal}. It
leads to smaller values of $\Delta_l$ than for the field parallel to the QR. Since the overlap factor $\Xi_l$ is
very weakly modified by the perpendicular field (Fig. 3b), the relaxation time given by Eq. \ref{t1_r} is reduced but still $T_1 \geq 1$s for $l>0$ is accessible (see Fig. \ref{f_delta_Ksi_T1_B_prost} and Table \ref{table3}). 

To build a qubit on singly occupied QR of small thickness one could in principle make use of the other (than $l=0$) Bohm--Aharonov minima of the dispersion relation $E(B_{\bot})$. However due to the strong decrease of $T_{1}$
with increasing the magnetic field (Eq. \ref{t1_r}) the relaxation times for such qubits would be much shorter.

\begin{table}[htp]
\caption{The relaxation time $T_1^l$ for different values of the orbital number $l$ and the magnetic filed $ B_{\bot} $. $r_0=14$nm and $ k=0.95$ have been assumed.}
\begin{indented}
\item[]\begin{tabular}{@{}cccccc}
\br
{\bf B} [T]	& \textbf{\emph{l}=0} & \textbf{\emph{l}=1}& \textbf{\emph{l}=2}& \textbf{\emph{l}=3}& \textbf{\emph{l}=4} \\ \mr
0.6	&1.789		& 14.035		& 26.390		& 34.093		& 37.754 \\ \mr
0.8	&0.38		& 3.209		& 6.110		& 7.933		& 8.809 \\ \mr
{\bf 1.0}	&{\bf 0.110}		& {\bf 1.013}		& {\bf 1.953}		&{\bf  2.549}		&{\bf  2.838} \\ \mr
1.2	&0.04		& 0.392		& 0.766		& 1.004		& 1.121 \\ \mr
1.4	&0.016		& 0.174		& 0.345		& 0.456		& 0.510 \\ \mr
1.6	&0.007		& 0.086		& 0.173		& 0.229		& 0.257 \\ \mr
1.8	&0.003		& 0.046		& 0.093		& 0.125		& 0.140 \\ \mr
2.0	&0.002		& 0.026		& 0.054		& 0.072		& 0.081 \\ \br
\end{tabular}
\label{table3}
\end{indented}
\end{table}

To compare the results for $B_{\|}$ and $B_{\bot}$ compare the 7$^{\rm th}$ row in Table \ref{table2} with the 3$^{\rm rd}$ row in Table \ref{table3} (bolded rows).\\
The above model considerations have been done for InGaAs/GaAs rings but the underlying physics is similar to other systems with somehow different set of parameters. 
In GaAs and GaAs/AlGaAs nanosystems the spin $g_s$ factor changes in a range $g_{s}\sim 0.2-0.4$ \cite{hans}. 
Assuming that material properties entering Eq. \ref{t1_r} are roughly the same as for InGaAs/GaAs and
 $N_e=1$ we obtain, e.g., for the ring $B$ made out of material with $g_{s} = 0.4$, $T_{1}\sim 6.4$s and for the ring $C$ (with $g_{s} = 0.4$), $T_{1}\sim 43$s. 
However, one has to stress that these very long relaxation times have been obtained taking into account only SO mediated interaction with piezoelectric phonons. 
Considering also other mechanisms of relaxation, (e.g., due to fluctuations of the electric and magnetic field, deformational phonons, multiphonon processes, and circuit noise) which we neglected in the above model calculations, can further limit the relaxation time.

The spin decoherence time $T_{2}$ for nanosystems made out of III-V semiconductors is limited by HFI as it was shown \cite{golo} that SOI does not lead to pure dephasing. 
Several strategies have been proposed to decrease
the randomness in the nuclear-spin system which can be useful also for QRs.
Dynamic nuclear polarization \cite{xu} and putting the the nuclear spins in a particular quantum state \cite{Giedke} are very promising.
 The estimated decoherence times are
 $T_{2} \sim 10 - 100 \mu $s \cite{xu,yao,bluhm} for the considered magnetic field. An alternative
approach is to use a quantum ring with holes instead of electrons. For a hole the
hyperfine coupling is expected to be much weaker than for an electron
because of the p-symmetry of the valence band \cite{grynch}. Recent
experiments have shown that hole spins remain coherent an order of magnitude
longer than electron spins \cite{brunner}.

Because of the detrimental effect of nuclear spins one can use different material. If QRs were made not of III-V semiconductors (with non-zero
nuclear spin) but of the group IV isotopes with zero nuclear spins, the coherence times should be
 longer because of the absence or very small (in isotopically not purified) hyperfine interaction. 
 As a result one could then get $T_{2} = 2T_{1}$, which is a relatively long time. 

Besides 'natural' semiconductors there exists another material having amazing capabilities for electronics. 
Carbon nanotubes constitute a new class of ballistic low dimensional quantum systems which also can be used for the implementation of a qubit \cite{Kuem,trau}. 
They are attractive because the zero nuclear spin of the dominant isotope $^{12}C$ yields a strongly reduced hyperfine interaction.

Decoherence times both for QDs and QRs depend strongly on the material used, however it seems that the reduced dimensionality of the device leads to an increase of decoherence times \cite{MHD,simon}. Because of ubiquitous nature of Si in modern electronics the estimations for Si rings \cite{lee} are important. 
It is known that the magnitude of the SOI in Si is ten times smaller than in GaAs and thus the relaxation times should be hundred times longer. 
However, for Si and SiGe systems $g_{s}\sim 2$ and these two factors make $T_{1}$
of the same order as for GaInAs rings. 
At the same time these systems should have long decoherence times $T_{2} = 2T_{1}$ due to the absence of nuclear spins. 
Si, the best semiconducting material for charge based electronics also seems to be a promising choice for spintronics and for quantum computing \cite{aws}. 
Summarizing, the decoherence times of electron spins in material with few or no nuclear spins as well as decoherence times for hole spins are expected to be much longer than for the group III-V semiconductors. 
However, in all considered materials the decoherence times are much longer than the initialization, qubit operations and measurement times allowing for quantum error correction scheme to be efficient. Recently a significant reduction of the randomness in the nuclear field reducing electron spin dephasing has been investigated both experimentally and theoretically \cite{vink, bucholz, giesbers, csonka}.

\section{Discussion and possible experimental realizations of quantum rings}\label{sec4}

The crucial point for quantum information processing is the necessity to keep coherence for a sufficiently long time. 
Based on the model calculations restoring roughly the energy spectra of experimentally feasible rings we have shown 
that quantum rings placed in a static magnetic field can be resistant to relaxation due to spin-orbit mediated
 electron-phonon interaction that is the main source of spin relaxation at magnetic fields $ 0.5 $T$ < B < 10$T.
It is known that QDs can have long relaxation times of the order of 1s at $B=1$T \cite{amasha,kroutvar}. Thus we asked the question whether such long $T_{1}$ can be reached also in QRs. The relaxation (and decoherence) time depends on the relevant orbital energy gap and the overlap factor which can be modified by changing the size and thickness of the ring. 
It also depends on the factor $g_s$, number of electrons, material parameters and different systems were examined to optimize coherence. 
The estimated relaxation times (at $B=1$T) for the experimentally fabricated rings are in the range between a few milliseconds to a few seconds.

Looking for qubits with relaxation times exceeding seconds ({\it $T_{1}$>1s}) our results can be summarized as follows:\\
a){\it for nanosystems with $r_{0}>12$nm the rings with $N_{e}>1$ are required (the advantage of energy gaps $\Delta_{l}$ increasing with $l$). }

Such rings can be produced by (i) methods relying on self assembled growth (SQRs), (ii) methods using nanolitographical procedures and electrostatic potentials (EQRs). 
Both methods have been successfully applied to obtain ring structures with one dimensional confinement of carriers (electrons and/or holes). Method (i) consists of modifications of growth procedures applied for quantum dot formations. 
After the Stranski-Krastanov (S-K) growth of QDs the surface is covered with an appropriate capping layer, and then the structure is annealed for the appropriate time. This method of transforming S-K QD into QR structures has been successfully applied to InGaAs QDs grown on GaAs surfaces by different epitaxial methods - metalorganic vapor phase epitaxy \cite{aierken} and molecular beam epitaxy \cite{bara,mlakar}.

Another method, namely liquid droplet epitaxy can be applied, if in the process of vapor phase epitaxial growth of binary, or multinary materials, one of the constituting elements can occur in liquid phase at the growth surface.
This method is usually used for growing QR like structures from GaAs and InGAs \cite{lee1, tong, zhao}.
Usually the SQRs structures obtained by the S-K QD growth followed by a post-growth annealing, as well as those grown by droplet epitaxy have radii in the range of 14 to 50 nm \cite{lei,kot,mano,Abba}.
In method (ii) the appropriate potential profiles confining carriers into QR geometry can be realized either by etching the QR like patterns from the layered heterostructures \cite{mohseni} or by depositing the metallic gates defining the appropriate electrostatic potential profiles, on top of structures with 2-dimensional electron (or hole) gas \cite{martensson}. \\
b) {\it QRs with $r_{0}<12$nm occupied by a \textit{single electron} have the required relaxation times.}

Despite the difficulties in producing such small structures the successful realization of the MBE grown InAs QR structures, with radius of 11.5nm has been reported \cite{kleemans} and used in our considerations. 
Another possibility of practical realization of the QR structures with radii in the range of 10 nm (or less) is the combination of axial and radial heterostructures in one semiconductor nanowire
(NW), i.e. the growth of a heterojunction in the NW shell with appropriate combination of materials, for example a NW with AlGaAs core, GaAs shell, and thin InAs section within the GaAs shell only. 
This type of NW structures have not been realized yet, to our knowledge, however due to the very rapid progress in the NW growth technology the future realization of such structures can be anticipated. 
Recently Mohseni et. al. \cite{mohseni} reported the QR-like confinement of electrons on top parts of core-shell GaAsP/GaP nanowires grown by molecular beam epitaxy.
 The use of nanowires has some advantages in comparison to the QD or QR structures fabricated by the S-K or liquid droplet epitaxy growth methods.
NWs can easily be grown on pre-patterned substrates, i.e. they can be oriented in plane in periodical structures, defined by the patterning process \cite{martensson}. 
The dimension of NWs can also be precisely controlled. In particular the radii of NWs can be smaller than 5 nm \cite{shtrikman}, which is rather impossible in case of self assembled QDs. 

The successful realization of structures which are closer to nano-ring geometries namely semiconductor nanotubes was also reported for some semiconducting materials like Si \cite{mbenkum}, ZnO \cite{kong}, or GaN \cite{goldberger}. However realization of QR structures would need implementation of axial heterostructures in nanotube geometries, which has not been demonstrated yet, to our knowledge. Another possibility is to define a set of QRs in a nanotube by local gate electrodes.
Probably the combination of confinement due to the electric potential defined by metallic gates in the small sections of core shell nanowire structures, already successfully applied to define QD like potential \cite{shtrikman} seems, at present to be the best method for realization of QR type confinement studied in our paper. 
It seems that the relatively thin rings of radius smaller than $10$nm with relaxation times exceeding seconds are within reach for nowadays nanotechnology. 
 

\section{Summary}\label{sec5}
Quantum information processing and spintronics have been major driving forces towards full control of single-spin systems. 
In particular fascinating phenomena based on carrier confinement in ring shaped nanostructures have intrigued physicists for many years. 
It was found that nanorings with $R< 20$nm can be considered as almost ideal quantum systems \cite{Abba} and thus can be, besides QDs, excellent systems for spin studies. 

We have investigated quantum rings with a single or a few electrons and have shown that they can be treated as quantum bits fulfilling DiVincenzo criteria \cite{divincenzo}. We have shown that for both QDs and QRs long relaxation times exceeding seconds at $B=1$T are possible.
It follows from our analysis that for singly occupied structures QDs have always longer relaxation time than QRs but for relatively thin rings with higher occupation the relaxation times can exceed those for QDs.
However (see e.g. Table \ref{table1}) even singly occupied rings with $r_0\leq 10$nm can have relaxation times exceeding seconds. 
The single occupancy of rings makes the experiments and the analysis more transparent, however, there is an open question whether qubits with $N_e>1$ can have some other advantages over those with $N_e=1$. %
 
The presented considerations demonstrate the feasibility of operating single-electron spin in QR as the quantum bit. This is of big relevance for the use in quantum information processing devices.

Finally, it should be stressed that multiply connected ring geometry offers additional (orbital) degree of freedom to be used for quantum manipulations. 
It is possible to build a qubit also on the orbital degrees of freedom \cite{szop,zipp} in some analogy to flux qubits on superconducting rings. 
Thus quantum carrier confinement in circular nanostructures can be the basis of many
applications in quantum information processing devices.

\ack
One of the authors (M.M.M.) acknowledges Grant No. NN 202$\:$128$\:$736 from
Ministry of Science and Higher Education (Poland). E.Z. thanks T. Dietl and W. Jantsch for valuable discussions.

\end{document}